\documentclass{article} \usepackage{epsf} \usepackage{amsmath}
\usepackage{bbm,mathrsfs}

\newcommand{\I}{\mathrm{i}}

\newcommand{\Prj}[1]{\mathrm{P}_{\scriptscriptstyle #1}}	
\newcommand{\LPrj}[1]{\mathcal{P}_{\scriptscriptstyle #1}}	

\newcommand{\One}{\ensuremath{I}}
\newcommand{\Eins}{\ensuremath{\mathbbm 1}}
\newcommand{\lattD}{\ensuremath{\mathscr{D}}}
\newcommand{\bmatr}{\ensuremath{\mathscr{E}}}

\newcommand{\btr}{\mathrm{Tr}\,}	

\newcommand{\C}{\mathrm{C}}
\newcommand{\Cbar}{\overline{\C}}

\newcommand{\DD}{\mathrm{D}}
\newcommand{\D}{\mathrm{d}}

\newcommand{\psibar}{\overline{\psi}}

\newcommand{\chibar}{\overline{\chi}}

\newcommand{\sprod}[2]{\langle #1,\, #2 \rangle}

\newcommand{\com}[2]{\left[ #1 , #2 \right]}
\newcommand{\anticom}[2]{\left\lbrace #1 , #2 \right\rbrace}

\newcommand{\slashed}{\displaystyle{\not}}

\begin{document}
\thispagestyle{empty} \parskip=12pt \raggedbottom

\noindent
\vspace*{1cm}
\begin{center}
  {\LARGE Towards Weyl fermions on the lattice without artefacts }

 
  \vspace{1cm} 
  Peter Hasenfratz and Reto von Allmen \\
  Institute for Theoretical Physics \\
  University of Bern \\
  Sidlerstrasse 5, CH-3012 Bern, Switzerland
  
  \vspace{0.5cm}
  
  \nopagebreak[4]
 
\begin{abstract}
In spite of the breakthrough in non-perturbative chiral gauge theories
during the last decade, the present formulation has stubborn
artefacts. Independently of the fermion representation one is
confronted with unwanted $CP$ violation and infinitely many
undetermined weight factors. Renormalization group identifies the
culprit. We demonstrate the procedure on Weyl fermions in a
real representation.
\end{abstract}

\end{center}

\vfill

\eject

\section{Introduction and Summary}

The non-perturbative formulation of chiral gauge theories has been
blocked for a long time by the problem of chiral symmetry in vector
theories. Although the Ginsparg-Wilson (GW) relation\cite{GW1982},
which is coding chiral symmetry on the lattice, was around since 1982,
it took a long time to appreciate its
significance\cite{Hasenfratz:1997ft}, find acceptable Dirac operators
satisfying the GW relation\cite{Neuberger:1997fp} and identify the
modified chiral 
symmetry transformation\cite{Luscher:1998pqa}. Rather different
approaches\cite{Kaplan:1992bt} 
converged to the conclusion that chiral vector theories (like QCD
with zero quark masses) can be defined non-perturbatively without
compromising any of the basic principles of a QFT. This development
lead to a formulation of chiral gauge theories on the lattice where it
became possible to demonstrate that the theory has exact gauge
symmetry and satisfies the basic requirements of a QFT in every order
in a perturbative expansion, or non-perturbatively\cite{Luscher:1999mt}.

The starting point in the works\cite{Luscher:1999mt} is a vector gauge
theory where the Dirac operator 
satisfies the GW relation. The fermions are chosen to be in a gauge
anomaly free representation of the target chiral gauge theory. The
next step is to introduce left-handed fermion fields. It turns out
that under the condition of locality the gauge field dependent
projectors\cite{Luscher:1998du} show an asymmetry between fermions and
anti-fermions. This is the source of fermion number violation in the
chiral theory, and so, it is a welcome feature. On the other hand, this
asymmetry breaks $CP$ at the same time\cite{Hasenfratz:2001bz}. The
r\^ole of this 
unwanted symmetry breaking in the chiral theory is not completely
understood. Since the generated $\theta$-term is a 4-dimensional
operator, this $CP$ violation creates a tuning problem. The number of
degrees of freedom depends on the topological sector which, on one
hand creates fermion number violation (as expected), on the other
hand, is responsible for the fact that the chiral theory falls into
topological sectors, where the relative weights between these sectors
remain undetermined.

For the problem discussed here, the basic step in the process above is the first one: choosing the
Dirac operator in the vector theory. Using
renormalization group (RG) language, which was the way to find the GW
relation in the first place\cite{GW1982}, we might define the lattice
action in the Gaussian fixed point\cite{Hasenfratz:1993sp}. For example, one might use
a simple ``blocking out of continuum''
step\cite{Bietenholz:1995cy}. We shall consider symmetries like the
axial and vector flavour transformations, where the gauge field is not
influenced. Concerning the
form of the block transformation one can make the following two
statements:
\begin{enumerate}
\item One {\em can} break any of such global symmetries in the block
  transformation if this is convenient.
\item One {\em must} break in the block transformation those
  symmetries which are anomalous in the quantized target theory.
\end{enumerate}

The first statement follows from the fact that the RG group
transformation in ref.\ \cite{Hasenfratz:1993sp} does not change the physical content of the
symmetries considered. The symmetries of the action broken by the blocking
do not disappear, rather the symmetry transformations  have a new
form\cite{Luscher:1998pqa}. Actually, there exists a simple and general procedure to
find these modified symmetry
transformations\cite{Hasenfratz:2006kv}. The modified symmetry
transformations might be gauge field dependent and the integration
measure of the path integral is then not necessarily invariant. The
symmetry is called anomalous if the measure cannot be kept invariant
without violating basic principles of a QFT.

The second statement is obvious: if a symmetry of the formal continuum
action is not broken by the blocking which brings it to the lattice,
then this symmetry and the continuum symmetry transformation will be
inherited by the lattice action and the measure. There will be no
anomaly and the punishment is non-locality, or unwanted particle
content (``doublers''), or something else. In our problem one might
introduce fermion number violation by hand in the projectors (rather
than via RG), but this step leads to the artefacts discussed before.

We shall demonstrate on a $SU(2)$ gauge theory with two flavours that
the RG indeed avoids the artefacts if statement 2.\ above is
respected. The lattice action of the vector theory is invariant under
the $U(1)$ axial and $U(1)$ vector flavour transformations which are
gauge field dependent. The correct axial anomaly is reproduced by the
measure, while the fermion number conservation remains intact in this
vector theory.
The vector theory falls into left- and
right-handed pieces with the gauge field independent projectors
$\Prj{R/L}=\frac{1}{2}(1 \pm \gamma_5)$. The chiral theory has the
expected $U(1)$ flavour anomaly breaking fermion number conservation.

This chiral gauge theory with Weyl
fermions has been discussed earlier by Suzuki\cite{Suzuki:2000ku} using the standard
setup. In spite of the real representation, the artefacts were present
there the same way as in theories in complex
representations\cite{Suzuki:2000ku}.

The difficulty to bring these ideas over to complex representations
lies in finding a block transformation which is theoretically
obviously correct and technically feasable at the same time.

\section{The lattice action with two GW type of relations}

We start with a formal vector $SU(2)$ gauge theory with two flavours
in the continuum. With a blocking out of continuum RG
step\footnote{Since finding the Gaussian fixed point is a classical
  field theory problem\cite{Hasenfratz:1997ft,Hasenfratz:1993sp}, the ``blocking out of continuum'' is
  equivalent to blocking the Wilson action on a lattice with lattice
  unit $a\rightarrow 0$ assuming that the averagings are matched.} we bring
the theory to the lattice. Since we consider a block transformation
with fermion number breaking (see statement 2.), it is useful to
introduce an 8-component notation in the form\footnote{We are indebted
  to Ferenc Niedermayer for suggesting this formalism.}
\begin{align}
  \label{eq:1}
  \phi (x) &= \begin{pmatrix} \psi (x) \\ \psibar^T (x) \end{pmatrix}.
\end{align}
In eq.~(\ref{eq:1}) the flavour ($i= 1,2$), colour ($a=1,2$) and the
Dirac ($\alpha = 1, \ldots, 4$) indices are not shown explicitly. The
fixed-point (FP) lattice fermion action is obtained by a simple
minimization (\cite{Hasenfratz:1997ft,Hasenfratz:1997nt,vonAllmen:2007} and references therein)
\begin{align}
  \label{eq:2}
  \dfrac{1}{2} \varphi^T \lattD \varphi &= \min_\phi \lbrace
  \dfrac{1}{2} \phi^T D
  \phi + \dfrac{1}{2}(\varphi - \Omega \phi )^T \bmatr (\varphi - \Omega \phi) \rbrace.
\end{align}
This equation defines the lattice fermion action (lhs.\ of
eq.~(\ref{eq:2})) for any given fermion configuration $\varphi$ on the
lattice by the rhs.\ of eq.~(\ref{eq:2}), where $\phi_{\min} =
\phi_{\min} (\varphi)$. In our notation
\begin{align}
  \label{eq:3}
  \varphi_n &= \begin{pmatrix} \chi_n \\ \chibar_n^T \end{pmatrix},
\end{align}
while $D$ and $\Omega$ are the continuum Dirac operator and the
averaging function (including the form of parallel transporting),
respectively. The continuum Dirac operator can be written as
\begin{align}
  \label{eq:4}
  D &= \begin{pmatrix} 0 & -d^T\\ d & 0 \end{pmatrix}, \quad d =
  \gamma_\mu D_\mu,
\end{align}
where $D_\mu$ is the covariant derivative and $d$ is diagonal in
flavour space\footnote{The path integral over the field $\phi$ with
  the action $\frac{1}{2}\phi^T D \phi$ gives the Pfaffian of $D$,
  $\text{Pf}\,(D)$. Here $\text{Pf}\,(D)^2 = \det D = (\det d)^2$,
  demonstrating that there is no double counting in the 8-component
  formulation.}. The gauge field ($U$) dependent averaging function has
the form
\begin{align}
  \label{eq:5}
  \Omega (U) &= \begin{pmatrix} \omega(U) & 0 \\ 0 & \omega^*(U)
  \end{pmatrix}, 
\end{align}
while the matrix $\bmatr$ is chosen to break the $U(1)$ chiral and $U(1)$
fermion number symmetries\footnote{In our conventions $\Cbar =
  \begin{pmatrix} \I \sigma^2 &0 \\ 0 & \I \sigma^2 \end{pmatrix}$.}
\begin{align}
  \label{eq:6}
  \bmatr &= \I \epsilon^{\text{c}} \cdot \epsilon^{\text{fl}} \cdot \Cbar \cdot \One, \quad \Cbar =
  -\Cbar^T = -\Cbar^\dagger,
\end{align}
where $\epsilon^{\text{c}}$ and $\epsilon^{\text{fl}}$ are the $2\times 2$
anti-symmetric $\epsilon$-tensors in colour and flavour space
respectively. The matrix $\One$ is the $8\times 8$ unit matrix in the
notation of eq.~(\ref{eq:1}). Due to the identity
\begin{align}
  \label{eq:7}
  \epsilon_{\xi \xi'} V^{\xi\eta} V^{\xi'\eta'} = \det V \cdot
  \epsilon_{\eta\eta'} = \epsilon_{\eta\eta'}
\end{align}
for $V \in SU(2)$, we find that the block transformation in
eq.~(\ref{eq:2}) is gauge invariant and preserves the lattice rotation
symmetries. It is also $C$, $P$, $T$ and $SU(2)$ flavour and $SU(2)$
chiral invariant. On the other hand, the block transformation breaks
$U(1)$ flavour (fermion number) and $U(1)$ chiral symmetries.

For any given lattice configuration $\varphi$ the corresponding
minimizing continuum field reads
\begin{align}
  \label{eq:8}
  \phi_{\min} (\varphi) = A^{-1} \Omega^T \bmatr \varphi, \quad A = D +
  \Omega^T \bmatr \Omega
\end{align}
which gives for the lattice Dirac operator
\begin{subequations}
\begin{align}
  \label{eq:9a}
  \lattD &= \bmatr - \bmatr \Omega A^{-1} \Omega^T \bmatr,
\end{align}
or, if $\lattD^{-1}$ exists,
\begin{align}
  \label{eq:9b}
  \lattD^{-1} &= \Omega D^{-1} \Omega^T + \bmatr^{-1}. 
\end{align}
\end{subequations}
In our 8-component notation the matrices related to flavour $U(1)$
axial and vector transformation in the continuum have the form
\begin{align}
  \label{eq:10}
  \Gamma_5 = \begin{pmatrix} \gamma_5 & 0 \\ 0 & \gamma_5
  \end{pmatrix}, \quad \Gamma_V = \begin{pmatrix} \Eins & 0\\ 0 &
    -\Eins \end{pmatrix}.
\end{align}
These matrices anti-commute with the continuum Dirac operator $D$
(eq.~(\ref{eq:4})) expressing $U(1)$ axial and vector (fermion number)
symmetries. The corresponding modified lattice symmetries of the
action are coded in two Ginsparg-Wilson type of relations
\begin{subequations}
\begin{align}
  \label{eq:11a}
  \anticom{ \Gamma}{ \bmatr^{-1} \lattD} &= 2 (\bmatr^{-1} \lattD) \Gamma
  (\bmatr^{-1} \lattD), 
\end{align}
where $\Gamma = \Gamma_5$, or $\Gamma = \Gamma_V$. If $\lattD^{-1}$ is
defined (no zero modes) then eq.~(\ref{eq:11a}) can be written as
\begin{align}\label{eq:11b}
\anticom{\Gamma}{\lattD^{-1}} &= 2\Gamma \bmatr^{-1}, \quad \Gamma=
  \Gamma_5 \text{ or } \Gamma_V.
\end{align}
\end{subequations}
The modified
infinitesimal lattice symmetry transformations read\cite{Hasenfratz:2006kv}
\begin{align}
  \label{eq:12}
  \delta\varphi &= \I \eta \Gamma ( 1 - \bmatr^{-1} \lattD ) \varphi,
  \quad \Gamma = \Gamma_5 \text{, or }  \Gamma_V.
\end{align}
With the help of the relations eq.~(\ref{eq:11a}) it is easy to show
that the lattice action $\frac{1}{2} \varphi^T \lattD \varphi$ is
invariant under the transformations in eq.~(\ref{eq:12}).

Switching off the gauge interaction, the Dirac operator $\lattD$ on
the lattice can be constructed explicitly and locality, and the
absence of doublers can be confirmed\cite{vonAllmen:2007}.

\section{The anomaly in the vector theory}
\label{sec:anomvec}

We show now that the measure of the vector theory 
\begin{align}
  \label{eq:13}
  \DD \varphi &\equiv \prod_{n,a,i,\alpha} \D \varphi_a (n)^{\alpha}_i  
\end{align}
is invariant under $U(1)$ flavour and anomalous under $U(1)$ axial
transformation, as expected\footnote{In eq.~(\ref{eq:13}) the indices
  $a, i, \alpha$ refer to colour, flavour and Dirac space,
  respectively.}.

The change of the measure under the $U(1)$ transformations in
eq.~(\ref{eq:12}) reads
\begin{align}
  \label{eq:14}
  \DD \varphi &\rightarrow \left( 1 - \I\eta \btr \left(\Gamma \bmatr^{-1} \lattD
    \right) \right) \DD \varphi, \quad \Gamma = \Gamma_5, \Gamma_V
\end{align}
where we used $\btr \Gamma =0$. We introduce an orthonormal basis to
calculate the trace above. The matrices $D$, $\bmatr$ and $\lattD$
are $\slashed{\Gamma}_5$-hermitian: $D^\dagger = \slashed{\Gamma_5} D
\slashed{\Gamma}_5, \ldots$, where
\begin{align}
  \label{eq:15}
  \slashed{\Gamma}_5 &=  \begin{pmatrix} 0 & \gamma_5\\ \gamma_5 & 0 \end{pmatrix}.
\end{align}
We consider the basis spanned by the eigenvectors $u_l$ of the
hermitian matrix $\hat{\lattD} = \slashed{\Gamma}_5 \lattD$. The
subspace of zero modes of $\hat{\lattD}$ is the same as that of
$\lattD$. Since $\lattD^{-1}$ is well defined on the non-zero modes,
we can write
\begin{align}
  -\btr \left( \Gamma \bmatr^{-1} \lattD \right) &= -
  \sum_{\lambda_l \neq 0} \sprod{u_l} {\Gamma \bmatr^{-1} \lattD
    u_l} = -\dfrac{1}{2} \sum_{\lambda_l \neq 0} \sprod{u_l} {(\Gamma
    \lattD^{-1} + \lattD^{-1}\Gamma) \lattD u_l} \nonumber \\
  &= \left \lbrace \begin{aligned} &0& \quad\text{if } \Gamma =
      \Gamma_V \\ 
      &-\sum_{\lambda_l \neq 0} \sprod{u_l}{\Gamma_5 u_l} =
      \sum_{\lambda_l = 0} \sprod{u_l}{\Gamma_5 u_l}, &\quad \text{if
      }\Gamma =\Gamma_5
    \end{aligned} \right. \label{eq:16}
\end{align}
where we used eq.~(\ref{eq:11b}) and the relations
\begin{align}
  \label{eq:17}
  \anticom{ \Gamma_V}{\slashed{\Gamma}_5} =0, \quad
  \com{\Gamma_5}{\slashed{\Gamma}_5} =0.
\end{align}
As eq.~(\ref{eq:16}) shows, the $U(1)$ vector symmetry is respected by
the measure, while only the zero modes
contribute to the axial anomaly.

From the GW relations in eq.~(\ref{eq:11a}) follows that if
$\hat{\lattD} u_0 =0$, then $u_0$,\newline $\frac{1}{2}(I \pm \Gamma) u_0$
($\Gamma = \Gamma_5, \Gamma_V$) and $\slashed{\Gamma}_5 u_0^*$ are
zero modes of $\lattD$. We can define then a convenient basis in the
space of the zero modes as
\begin{align}  \label{eq:cm18} 
  u_i = \begin{pmatrix} v_i \\ 0 \end{pmatrix}, \quad \bar{u}_i =
  \begin{pmatrix} 0 \\ v_i^* \end{pmatrix} \quad i = 1,\ldots,N.
\end{align}
Here $v_i$ and $v_i^*$ can be taken left- or right-handed, $v_{Li}=
\frac{1 -\gamma_5}{2} v_i$, $v_{Ri}=
\frac{1 +\gamma_5}{2} v_i$, $v^*_{Li}=
\frac{1 +\gamma_5}{2} v^*_i$, $v^*_{Ri}=
\frac{1 -\gamma_5}{2} v^*_i$. We have then $N_L$, $N_R$, $\bar{N}_L$,
and $\bar{N}_R$ zero modes where
\begin{align}
  \label{eq:19}
  \bar{N}_R = N_L, \quad \bar{N}_L = N_R, \quad N_R + N_L =N.
\end{align}
We obtain for the measure contribution
\begin{align}
  \label{eq:20}
   -\btr \left( \Gamma_5 \bmatr^{-1} \lattD \right) &= \sum_i \left(
     \sprod{u_i}{\Gamma_5 u_i} + \sprod{\bar{u}_i}{\Gamma_5 \bar{u}_i}
   \right) \nonumber\\
   &= N_R + \bar{N}_L - N_L - \bar{N}_R = 2(N_R - N_L).
\end{align}

The result $\I\eta\cdot 2 (N_R - N_L)$ for the change of the measure under
an infinitesimal $U(1)$ axial transformation is independent of the
details of the GW relation and is reproduced correctly in
eq.~(\ref{eq:20}).

\section{Fermion number anomaly in the chiral theory}
\label{sec:ferm-numb-anom}

Since both the continuum action and the block transformation in
eq.~(\ref{eq:2}) fall in left- and right-handed pieces with the
standard $\frac{1 \pm \gamma_5}{2}$ projectors, so does the lattice
action. In the 8-component notation the chiral field is written as
\begin{align}
  \label{eq:21}
  \phi_L &= \begin{pmatrix} \psi_L \\ \psibar_L^T \end{pmatrix} =
  \LPrj{L} \phi.
\end{align}
The infinitesimal fermion number transformation in the continuum reads
\begin{align}
  \label{eq:22}
  \delta \phi_L = \I \eta \Gamma_V \phi_L = -\I \eta \Gamma_5 \phi_L.
\end{align}
The corresponding transformation on the lattice has the form
\begin{align}
  \label{eq:23}
  \delta \varphi_L &= -\I \eta \Gamma_5 ( 1 - \bmatr^{-1} \lattD) \LPrj{L}
  \varphi_L
\end{align}
and the measure is changed by
\begin{align}
  \label{eq:24}
  -\text{Tr}_L\, \Gamma_5 (1 - \bmatr^{-1} \lattD).
\end{align}
The trace in eq.~(\ref{eq:24}) is taken in the left-handed zero mode
space only. One obtains from eq.~(\ref{eq:20})
\begin{align}
  \label{eq:9}
  -\text{Tr}_L\,\Gamma_5 (1 - \bmatr^{-1} \lattD) &= \bar{N}_L - N_L
\end{align}
which shows that the fermion number is anomalous in the chiral theory.

{\bf Acknowledgements}
The authors are indebted for discussions with Moritz Bissegger, Ferenc
Niedermayer and Uwe-Jens Wiese.\\
This work was supported by the Schweizerischer Nationalfonds.

\eject


\end{document}